# scientific **data**

**OPEN**

**DATA DESCRIPTOR**

# Dataset combining EEG, eye-tracking, and high-speed video for ocular activity analysis across BCI paradigms

Eva Guttmann-Flury ✉, Xinjun Sheng & Xiangyang Zhu

In Brain-Computer Interface (BCI) research, the detailed study of blinks is crucial. They can be considered as noise, affecting the efficiency and accuracy of decoding users' cognitive states and intentions, or as potential features, providing valuable insights into users' behavior and interaction patterns. We introduce a large dataset capturing electroencephalogram (EEG) signals, eye-tracking, high-speed camera recordings, as well as subjects' mental states and characteristics, to provide a multifactor analysis of eye-related movements. Four paradigms - motor imagery, motor execution, steady-state visually evoked potentials, and P300 spellers - are selected due to their capacity to evoke various sensory-motor responses and potential influence on ocular activity. This online-available dataset contains over 46 hours of data from 31 subjects across 63 sessions, totaling 2520 trials for each of the first three paradigms, and 5670 for P300. This multimodal and multi-paradigms dataset is expected to allow the development of algorithms capable of efficiently handling eye-induced artifacts and enhancing task-specific classification. Furthermore, it offers the opportunity to evaluate the cross-paradigm robustness involving the same participants.

## Background & Summary

Decoding the intricacies of the human brain stands as one of the paramount challenges in modern science, crucial for advancing neurorehabilitation, neuromorphic devices, and artificial intelligence. At the forefront of this endeavor lies Brain-Computer Interface (BCI) technology, which enables the translation of neural activity into actionable commands or insights. BCI systems require identifying neural patterns linked to mental processes, such as motor control, cognition, emotion, and perception[1,2].

Electroencephalography (EEG), with its non-invasive nature, accessibility, and high temporal resolution, is key for capturing the brain's electrical activity and studying neural dynamics in real-time[3]. Nonetheless, EEG recordings indistinctly integrate signals issued from all sources inside the brain. This makes them inherently susceptible to various sources of artifacts, notably those arising from ocular activity such as blinks or eye movements[4,5].

Voluntary blinks can serve as control commands in Human-Computer Interaction (HCI) applications, enabling communication through blink patterns[6]. However, spontaneous ones, pose a notable challenge in BCI by disrupting EEG potentials due to their rapid occurrence, lasting about $0.2 \pm 0.026$ seconds (min $= 0.044$ s, max $= 0.421$ s) with an average frequency of 20 per minute, as summarized in Table 3. This interference can obscure neural signals for over 0.3 seconds, constituting approximately 10% of a one-minute epoch[7]. Ocular movements like saccades and eye movements also produce artifacts, requiring careful preprocessing to ensure EEG-based findings' integrity and reliability[8].

The online availability of EEG BCI datasets has steadily increased[9], with platforms such as https://open-bci.com/community/publicly-available-eeg-datasets/ or https://physionet.org/about/database/. However, many datasets focus solely on blink recordings with restricted electrode coverage or specific BCI paradigms, limiting

State Key Laboratory of Mechanical System and Vibration, School of Mechanical Engineering, Shanghai Jiao Tong University, 800 Dongchuan Road, Minhang District, Shanghai, 200240, P. R. China. ✉e-mail: eva.guttmann.flury@gmail.com

| Subjects | Age | Gender | Height | Head circumference (cm) | Nasion - Inion (cm) | Eye correction Left (K dioptre) | Eye correction Right (K dioptre) | Laterality Index Handedness | Decile Edimburg Handedness | Augmented (15 item) index | Mother tongue | Familiarity with fast displays | Familiarity with BCI |
|---|---|---|---|---|---|---|---|---|---|---|---|---|---|
| S01 | 30 | F | 176 | 56 | 37 | 500 | 550 | 75 | 4th R. | 76.67 | FR | Very | Yes |
| S02 | 30 | M | 176 | 57 | 38 | 250 | 230 | 95 | 9th R. | 90 | FR | A little | Yes |
| S03 | 25 | M | 181 | 58 | 36 | 200 | 200 | 100 | 10th R. | 100 | ES | Very | No |
| S04 | 25 | M | 170 | 57.5 | 37 | 600 | 600 | -30 | 1st L. | -26.67 | ZH | A little | No |
| S05 | 23 | F | 163 | 56 | 35 | 850 | 900 | 80 | 5th R. | 86.67 | ZH | No | Yes |
| S06 | 25 | F | 168 | 57 | 33 | 0 | 0 | 25 | Middle | 33.33 | ZH | Very | No |
| S07 | 20 | M | 165 | 58 | 36 | 200 | 150 | 60 | 2nd R. | 66.67 | ZH | Very | No |
| S08 | 38 | F | 160 | 53 | 31 | 500 | 500 | 90 | 7th R. | 73.33 | ZH | Very | No |
| S09 | 29 | M | 185 | 59 | 38 | 175 | 150 | 75 | 4th R. | 76.67 | FA | Very | No |
| S10 | 28 | M | 186 | 56 | 37 | 200 | 100 | 100 | 10th R. | 100 | FR | Very | No |
| S11 | 29 | F | 170 | 54 | 33 | 100 | 100 | 75 | 4th R. | 76.67 | ZH | A little | Yes |
| S12 | 24 | F | 165 | 55 | 35 | 75 | 50 | 85 | 6th R. | 83.33 | RU | Very | No |
| S13 | 23 | M | 179 | 57 | 37 | 425 | 375 | -40 | 1st L. | -46.67 | ZH | A little | Yes |
| S14 | 27 | M | 180 | 61 | 39 | 300 | 300 | 95 | 9th R. | 96.67 | ZH | A little | Yes |
| S15 | 24 | F | 170 | 53 | 34 | 750 | 750 | 90 | 7th R. | 93.33 | ZH | Very | No |
| S16 | 22 | M | 172 | 60 | 34 | 900 | 800 | 95 | 9th R. | 90 | ZH | Very | No |
| S17 | 24 | M | 186 | 61 | 39 | 400 | 400 | -25 | Middle | -26.67 | ZH | A little | Yes |
| S18 | 26 | M | 182 | 57 | 37 | 175 | 175 | 80 | 5th R. | 86.67 | EN | Very | No |
| S19 | 32 | M | 180 | 57 | 36 | 0 | 0 | 55 | 1st R. | 50 | FR | Very | No |
| S20 | 27 | F | 159 | 56 | 33 | 350 | 350 | 100 | 10th D. | 96.67 | ZH | A little | No |
| S21 | 36 | F | 178 | 55 | 38 | 100 | 100 | 40 | Middle | 33.33 | FR | A little | No |
| S22 | 35 | M | 184 | 57 | 34 | 0 | 0 | 75 | 4th R. | 70 | FR | A little | No |
| S23 | 23 | M | 182 | 57 | 37 | 550 | 550 | 100 | 10th R. | 100 | FR | A little | No |
| S24 | 30 | M | 178 | 59 | 35 | 800 | 800 | 100 | 10th R. | 100 | ZH | Very | Yes |
| S25 | 26 | M | 158 | 59 | 38 | 0 | -200 | 80 | 5th R. | 86.67 | FR | Very | No |
| S26 | 33 | M | 182 | 57 | 38 | 0 | 0 | 100 | 10th R. | 100 | FR | Very | No |
| S27 | 25 | M | 175 | 57 | 33 | 300 | 200 | 35 | Middle | 23.33 | ZH | Very | Yes |
| S28 | 32 | F | 160 | 54 | 34 | NA | NA | 100 | 10th R. | 96.67 | FR | No | No |
| S29 | 32 | M | 168 | 57 | 36 | 0 | 0 | 65 | 2nd R. | 73.33 | FR | Very | No |
| S30 | 57 | F | 170 | 56 | 34 | NA | NA | 50 | 1st R. | 60 | FR | A little | No |
| S31 | 28 | M | 186 | 57 | 36 | NA | NA | 80 | 5th R. | 80 | EN & FR | A little | No |

**Table 1.** Demographic description and a few physical characteristics of the participants; R. for right, L. for left, D. for decile.

| Time (in s) | FP1 (in μV) | FPZ (in μV) | ... | Trig | Cues | PhanFrame | PhanTime (time stamp) | Blink |
|---|---|---|---|---|---|---|---|---|
| 24.521 | 17.19 | 14.69 | ... | 1 | Fixation | 166 | 09:33:49.145 | 1 |
| 24.527 | 18.34 | 16.42 | ... | 1 | Fixation | 167 | 09:33:49.151 | 0 |
| ... | | | | | | | | |
| 26.521 | 21.46 | 20.45 | ... | 1 | Right | 500 | 09:33:51.311 | 0 |

**Table 2.** Example of file merging the information from E-Prime, EEG, and high-speed video recordings.

insights into blink characteristics[10]. Additionally, most of these datasets have small sample sizes and few trials, raising concerns about overfitting and compromising the reliability and reproducibility of BCI research[11,12].

To enhance understanding of eye-related activities' impact on EEG data, we present a multimodal dataset combining EEG with eye-tracking and high-speed video. This integration allows precise identification of eye movements (identified with eye-tracking) and blinks (captured through video and EEG), leading to a better awareness of subjects' intra- and inter-variability. Beyond mere noise, eye movements signal cognitive states, fatigue, and attention, providing insights that enhance BCI design by clarifying how ocular activity affects neural signal interpretation.

Recording diverse BCI paradigms captures various cognitive and motor brain activities, each with unique signal processing challenges, such as varying signal-to-noise ratios and artifact types. Such diversity fosters the development of adaptable algorithms, enhancing robustness and applicability across different tasks and artifact types.

| Subject | Peak potential on FP1 Mean ± SD (in μV) | Blink frequency Mean ± SD (in blinks/min) | Number of blinks during tasks Mean ± SD (in %) | Blink width on FP1 Mean ± SD (in sec) | Blink width on FP1 Min – Max (in sec) |
|---|---|---|---|---|---|
| S01 | 220.4 ± 50.8 | 16.9 ± 5.0 | 17.1 ± 14.2 | 0.228 ± 0.015 | 0.056 – 0.393 |
| S02 | 158.3 ± 76.8 | 9.3 ± 1.9 | 10.3 ± 7.4 | 0.186 ± 0.02 | 0.048 – 0.418 |
| S03 | 115.0 ± 53.3 | 9.4 ± 2.5 | 7.3 ± 6.3 | 0.171 ± 0.029 | 0.044 – 0.387 |
| S04 | 214.4 ± 67.1 | 34.5 ± 7.8 | 31.3 ± 16.7 | 0.228 ± 0.055 | 0.046 – 0.449 |
| S05 | 250.3 ± 88.7 | 9.2 ± 4.3 | 9.7 ± 7.3 | 0.258 ± 0.037 | 0.047 – 0.484 |
| S06 | 142.7 ± 55.9 | 20.7 ± 3.4 | 20.1 ± 9.5 | 0.099 ± 0.012 | 0.034 – 0.288 |
| S07 | 144.3 ± 72.1 | 12.6 ± 2.9 | 15.5 ± 9.2 | 0.232 ± 0.03 | 0.053 – 0.485 |
| S08 | 211.4 ± 74.5 | 45.0 ± 14.4 | 49 ± 15 | 0.243 ± 0.079 | 0.041 – 0.515 |
| S09 | 195.5 ± 77.9 | 17.9 ± 5.0 | 20.5 ± 7.6 | 0.212 ± 0.029 | 0.047 – 0.493 |
| S10 | 178.6 ± 96.9 | 14.0 ± 3.5 | 11.8 ± 8 | 0.204 ± 0.026 | 0.046 – 0.464 |
| S11 | 279.0 ± 101.8 | 32.9 ± 5.9 | 31.5 ± 8.4 | 0.279 ± 0.044 | 0.041 – 0.567 |
| S12 | 142.5 ± 63.2 | 11.9 ± 3.3 | 10.7 ± 10.7 | 0.182 ± 0.023 | 0.045 – 0.398 |
| S13 | 127.7 ± 53.9 | 54.9 ± 10.9 | 45.9 ± 17.1 | 0.214 ± 0.02 | 0.047 – 0.474 |
| S14 | 127.7 ± 47.0 | 22.3 ± 4.3 | 19.5 ± 18.3 | 0.193 ± 0.033 | 0.044 – 0.409 |
| S15 | 201.4 ± 82.9 | 28.8 ± 6.2 | 32.6 ± 11.6 | 0.214 ± 0.012 | 0.039 – 0.466 |
| S16 | 135.9 ± 53.9 | 26.7 ± 8.5 | 24.8 ± 17.2 | 0.195 ± 0.041 | 0.036 – 0.484 |
| S17 | 106.6 ± 50.3 | 7.2 ± 1.5 | 5.3 ± 7.4 | 0.133 ± 0.028 | 0.041 – 0.3 |
| S18 | 155.6 ± 72.8 | 12.2 ± 4.2 | 12.2 ± 9.4 | 0.2 ± 0.019 | 0.046 – 0.475 |
| S19 | 105.2 ± 33.1 | 9.5 ± 1.2 | 8.7 ± 6.3 | 0.147 ± 0.01 | 0.044 – 0.315 |
| S20 | 306.1 ± 93.8 | 42.8 ± 3.5 | 33.7 ± 21.5 | 0.211 ± 0.021 | 0.034 – 0.459 |
| S21 | 178.3 ± 71.2 | 18.1 ± 2.5 | 19.3 ± 6.9 | 0.134 ± 0.043 | 0.032 – 0.389 |
| S22 | 94.5 ± 30.4 | 16.8 ± 2.0 | 19.3 ± 11 | 0.175 ± 0.005 | 0.045 – 0.375 |
| S23 | 101.6 ± 59.7 | 7.3 ± 1.6 | 6.3 ± 5.5 | 0.131 ± 0.022 | 0.04 – 0.326 |
| S24 | 142.3 ± 52.7 | 46.1 ± 3.4 | 39.9 ± 8.8 | 0.239 ± 0.02 | 0.046 – 0.48 |
| S25 | 85.7 ± 32.1 | 15.4 ± 4.2 | 16.9 ± 12.6 | 0.167 ± 0.038 | 0.039 – 0.397 |
| S26 | 100.2 ± 31.9 | 8.2 ± 2.4 | 9.8 ± 9.5 | 0.143 ± 0.022 | 0.043 – 0.347 |
| S27 | 142.2 ± 61.7 | 27.7 ± 4.5 | 21.5 ± 16.1 | 0.211 ± 0.013 | 0.046 – 0.433 |
| S28 | 199.8 ± 87.2 | 25.5 ± 1.7 | 23.9 ± 3.8 | 0.214 ± 0.007 | 0.042 – 0.433 |
| S29 | 83.8 ± 27.6 | 15.3 ± 3.4 | 14.6 ± 18.3 | 0.208 ± 0.017 | 0.042 – 0.307 |
| S30 | 121.5 ± 47.5 | 13.4 ± 2.2 | 17.1 ± 11.5 | 0.216 ± 0.015 | 0.048 – 0.448 |
| S31 | 194.7 ± 69.4 | 13.0 ± 4.2 | 11.6 ± 7.7 | 0.216 ± 0.015 | 0.053 – 0.38 |
| **Mean** | **160.1 ± 56.4** | **20.8 ± 12.8** | **19.93 ± 11.42** | **0.197 ± 0.041** | **0.044 – 0.421** |
| **SD** | **62.5 ± 20.4** | **4.3 ± 2.9** | **10.99 ± 4.53** | **0.026 ± 0.016** | **0.005 – 0.069** |

**Table 3.** Blink potential distribution parameters per subject.

........................................................................................................................................................

This study presents a multimodal dataset featuring 31 participants, both left and right-handed individuals, over 63 sessions. It includes 2520 instances each of Motor Imagery (MI), Motor Execution (ME), and Steady-State Visual Evoked Potentials (SSVEP), along with 5670 instances of P300 signals. To avoid voluntary blinking, participants were briefed on the BCI paradigm objectives without mentioning blinks. Demographic, physiological, and psychophysiological state assessment data were also gathered using questionnaires and biometric measurements (e.g., facial landmarks derived from photographs). Sample size calculations were performed for both task- and blink-related signals to ensure sufficient data coverage.

To the best of our knowledge, this large multimodal dataset uniquely provides simultaneous electrophysiological recordings, video capture, and synchronized eye-tracking, with all data and code available online for reproducing the experiment. Our goal is to enable the creation of signal processing algorithms that efficiently counteract eye-related artifacts and improve the accuracy of task-specific neural decoding. Additionally, we anticipate its utility in assessing BCI technique adaptability across various paradigms, advancing our understanding of cognitive processes and behavior.

This dataset provides a visually observable reference for eye movements, specifically blinks, through video recordings. This resource can facilitate the systematic evaluation of methodologies for EEG and eye-tracking, particularly in refining artifact correction algorithms. It also enables investigations of interactions between eye movements, EEG signals, and paradigms, advancing our understanding of their mutual influence. Furthermore, the dataset supports the development of algorithms to accurately identify paradigms, discriminate classes, and assess robustness across different scenarios. It supports investigations between self-reported cognitive states and objective measures, as well as research on blink biometrics and psychophysiological states. Overall, blinks provide frequent and detectable signals across subjects, offering rich opportunities for extracting meaningful information in EEG-based BCI research.

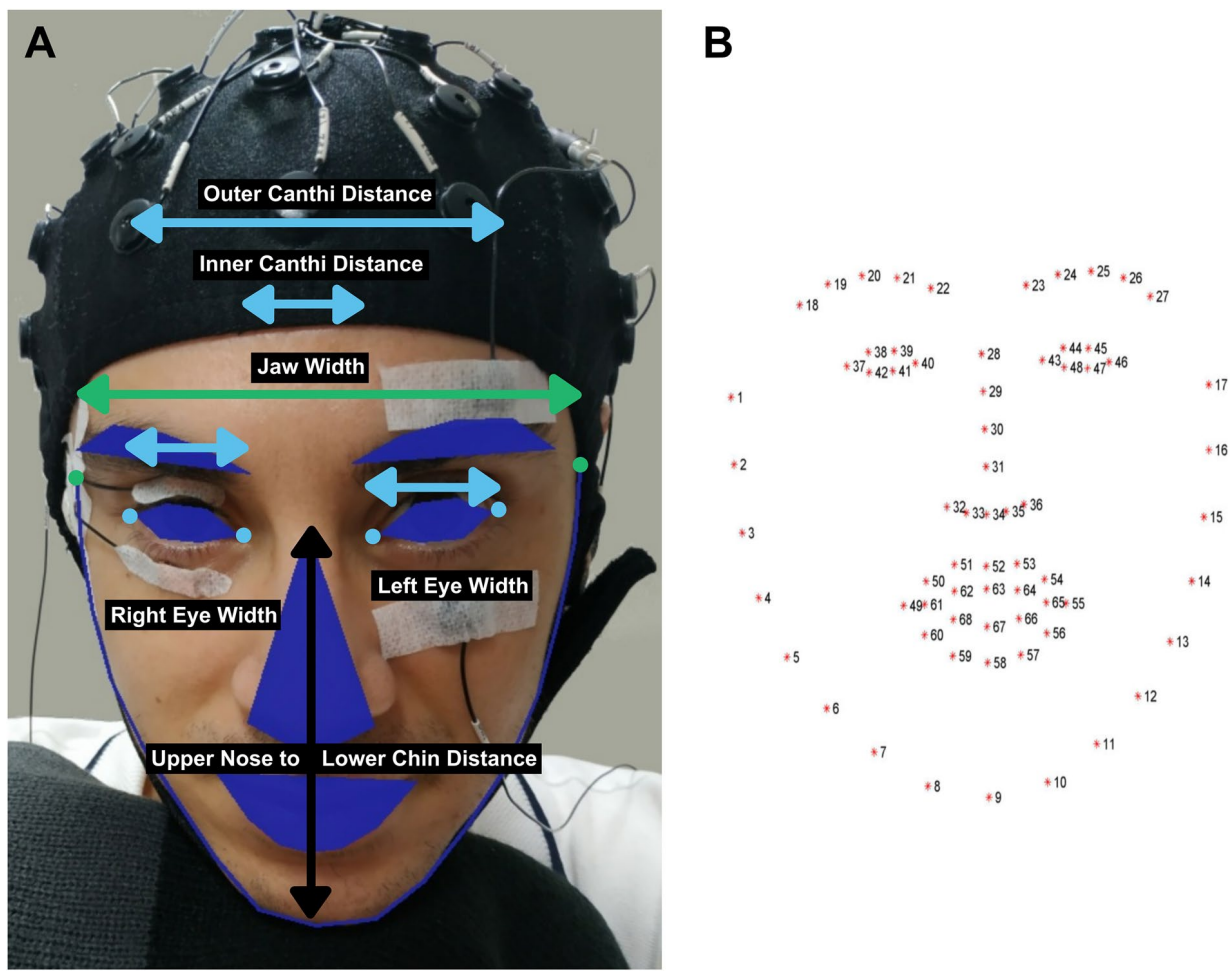

**Fig. 1** (**A**) Five subject-specific distances computed from (**B**) the 68 points iBUG 300-W dataset defined face regions. Informed consent was obtained from the individual in the figure for the publication of the images.

## Methods

**A priori sample size estimation.** The first step in any study seeking to replicate or build upon previous research involves analyzing the impact and sample sizes of prior investigations on the subject. The specific objectives of the research guide the calculation of required sample sizes, significantly influencing the experiment's design. For example, research focused on creating predictive models will necessitate varying sample sizes based on the nature of the variables being tested—be it binary, multivariate, or other types. This step is essential to ensure that the investigation has enough statistical power to identify significant effects or confirm the precision of the predictive models[13].

The focus of the presented dataset is twofold: blink analysis and BCI paradigm classification. Either the signals associated with blinks (e.g., blink maximum potential) or the cortical sources activated by a paradigm (e.g., distance from individual trial covariance matrix to class-averaged covariance matrix) can be investigated.

Once a signal of interest has been determined, data can be acquired from historical datasets or through prospective power analysis. The latter involves initially testing the experiment on a small subset of subjects and assumes that the population distribution reflects that observed in the initial sample. Experimental data can be fitted to a distribution, which in turn is used to generate simulated data with defined effect sizes. Monte Carlo simulation methods, used if the distribution is not normal, help determine the minimum sample size required to achieve an 80% power target based on the chosen effect size. Further details on executing this analysis, particularly using blink maximum potential from historical data, are elaborated in[14].

Consider an example where the experiment aims to simply detect blinks from EEG signals. With blink potentials averaging at $160 \pm 50\,\mu V$, significantly higher than other EEG data averaging at $30 \pm 20\,\mu V$, the Cohen's d effect size exceeds 3. Consequently, in a Monte Carlo simulation with a fitted distribution that follows the blink distribution, the required a priori sample size would certainly be very small, probably less than a minute of recording, in line with blinks occurring approximately 20 times per minute.

Now, let's explore another scenario focusing on variations in blinking across subjects. Here, the expected effect size is smaller, around 0.2, necessitating longer recording durations, typically spanning at least 45 hours[14]. Similar calculations can determine the number of sessions required for a specific paradigm. For example, if the signal of interest linked to Motor Imagery (MI) is considered, a prospective power analysis suggests a lower effect size of around 0.1, necessitating a minimum of 63 sessions in a Monte Carlo simulation, assuming standard alpha (0.05) and power levels (0.8).

When confronted with the choice among various potential sample sizes, the standard procedure is to opt for the larger value. For the current dataset, this implies recording a minimum of 63 sessions, which roughly corresponds to 46 hours of recording.

**Participants.** This study was approved by the Institutional Review Board of Shanghai Jiao Tong University, Protocol No. (IRB HRP E2021216I), date of approval (March 4th, 2021) and enrolled 31 healthy volunteers (11 women, 20 men; mean age 29 ± 7), who provided written consent for anonymized data use. A total of 63 sessions were recorded, with 14 participants completing one, 2 attending two, and 15 completing three. During the initial session, participants completed a questionnaire covering demographic, physiological, and psychophysiological state assessments[15,16]. Handedness or familiarity with BCI did not affect participation eligibility. They are designated using their aliases for anonymity (S01-S31), with relevant characteristics summarized in Table 1.

Individual photos were taken, and facial landmarks extracted using a detector trained on the iBUG 300-W dataset[17]. Seven face regions were defined: jaw, mouth, nose, left and right eye, and eyebrow. Pixel-based measurements were converted to metrics for consistency across photos, yielding five subject-specific measures: Left and Right Eye Width, Inner and Outer Canthi Distance, and Upper Nose to Lower Chin Distance, as illustrated in Fig. 1.

**Paradigms.** The participants all completed four BCI tasks: Motor Imagery (MI), Motor Execution (ME), Steady-state visual evoked potentials (SSVEP), and P300 visual evoked potentials. Each task was limited to 14 minutes to manage the high-speed camera storage and reduce user fatigue. The P300 speller task was split into two parts, one for four-letter words (P3004L) and the other for five-letter words (P3005L), totaling 31 minutes. Sessions were on average 45 minutes long, excluding breaks and questionnaires, and included all five tasks randomly ordered. The P3005L paradigm consisted of 50 trials, while the other tasks had 40 trials.

All tasks followed a similar structure. A greeting message appeared on screen, during which a 300 trigger code was sent along with a Welcome cue. Then instructions specific to the task were displayed, until a thank-you message app at the end, during which a 300 trigger code was sent along with a Goodbye cue. Trial durations were fixed, but Welcome and Goodbye recording times varied based on when the experimenter launched and stopped the Python code.

*Paradigm #1: Motor Imagery (MI of grasping with all fingers).* Participants are directed to engage in kinesthetic motor imagery by imagining grasping with either their left or right hand, involving all fingers. The 40 trials are evenly split between 20 Left Motor Imagery (MI) and 20 Right MI tasks. Each trial begins with a white fixation cross displayed at the screen's center for 2 seconds. Subsequently, a red rectangle cue randomly appears on either side of the cross for 4 seconds. Upon cue onset, participants initiate the mental task, imagining grasping the corresponding hand three times at a self-paced rhythm of approximately 1 Hz.

Following the task, the trial concludes with the disappearance of the fixation cross and red rectangle cue, transitioning into a random relaxation period lasting 1-1.5 seconds. This intertrial interval allows for relaxation while preventing subject adaptation. Trigger codes indicate the trial number, while cues are recorded as "Fixation," "Left," or "Right," along with "Break Random" (refer to Fig. 2).

*Paradigm #2: Motor Execution (ME of grasping with all fingers).* The motor execution (ME) experimental paradigm mirrors the one for motor imagery (MI), as depicted in Fig. 2. Trigger codes and cues remain consistent across both tasks. The sole distinction lies in participants physically executing the hand grasping movements during ME tasks.

*Paradigm #3: Steady-State Visual Evoked Potentials (SSVEP).* The stimuli consist of four black-and-white checkerboards positioned in each quadrant of the monitor, each flickering at a fixed frequency (10 Hz, 13 Hz, 12 Hz, and 11 Hz, respectively). The 40 trials are randomly arranged, with an equal distribution across each target frequency. Each trial begins with a 2-second presentation of a red arrow indicating the gaze direction, followed by a brief 0.5-second black screen interval.

Participants then fixate on the target stimulus for 4.5 seconds before a random rest period of 1-1.5 seconds with a black screen. The trigger code corresponds to the trial number, while the cues are labeled as "Stimulus", "Break", "F Hz" (with F representing either 10, 11, 12, or 13), and "Break Random" (see Fig. 2).

*Paradigm #4: P3004L (P300 for Four Letters Word).* Stimulation Sequence. In the conventional visual oddball paradigm, the identification of the P3b component is used to deduce the intended stimulus. A prominent example is the Farwell and Donchin speller, which features a matrix with cells that alternate in flashing. A sequence concludes once all cells in the matrix have been highlighted. Typically, each image reveals six symbols, and six such images form a sequence. A trial usually consists of three consecutive sequences. Originally, the six cells are organized based on the row/column paradigm (RCP), where either a complete row or column is flashed[18]. However, the checkerboard paradigm (CBP) has demonstrated notable performance improvements over the RCP by eliminating adjacent letters in vertical or horizontal orientation[19,20].

In our experiment, we employed the traditional white/gray flicker matrix containing the 26 letters of the Latin alphabet followed by Arabic numerals from 1 to 9 and the hyphen-minus symbol. For each trial, three sequences are randomly selected from 120 sequences generated using the CBP principle. Once all 36 symbols have been divided into six groups, this algorithm is repeated 120 times. Consequently, this algorithm can produce approximately 250000 compatible images (from $C_{36}^6 \approx 2000000$ images including adjacent characters) yielding around 4000 CBP sequences. A sequence is constructed using the following Algorithm 1, which generates six valid images.

**Algorithm 1.** Pseudo-code to generate CBP sequence for BCI P300 speller.

**Data:** a list of 36 characters C; AD list of horizontal and vertical adjacent characters: $AS(c_i) = c_{i-6}, c_{i-1}, c_i, c_{i+1}, c_{i+6}$ if they exist: $c_{i-6}$ if $i > 6$, $c_{i-1}$ if $modulo(i, 6) > 1$, $c_{i+1}$ if $modulo(i, 6) > 0$, $c_{i+6}$ if $i < 31$

**Result:** a sequence S of six images, each highlighting six characters

```
1  Initialize sequence and list of available characters: S = {}; AC = C
2  for i = 1 to 6 do
3  |   ACI = AC; J = {}; FCI = {};              // Initialize available, chosen, and forbidden
   |                                            //    characters for the current image
4  |   for j = 1 to 6 do
5  |   |   if FCI ∩ AC = ∅                       // No forbidden character should be in the available
   |   |                                        //    ones
6  |   |   then
7  |   |   |   c = random(ACI)                    // Randomly choose a symbol
8  |   |   |   IM = IM ∪ {c}                       // Saving chosen symbol
9  |   |   |   ACI = ACI − AD(c)                   // Remove adjacent symbols from available ones
10 |   |   |   AC = AC − c                          // Remove chosen character from available ones for all
   |   |   |                                        //    remaining images
11 |   |   |   FCI = FCI ∪ AD(c)                   // Saving forbidden characters
12 |   |   |   if j = 6 then
13 |   |   |   |   S = S ∪ {IM}                      // Save image containing six characters
14 |   |   |   end
15 |   |   else
16 |   |   |   break                               // Break when not enough remaining symbols
17 |   |   end
18 |   end
19 end
20 if card(S) = 6 then
21 |   return S
22 end
```

Cue and Trigger. Subjects are tasked with spelling words by focusing on the color change of each letter, initially gray against a dark background. To indicate the target letter, a green ellipse briefly encircles it at the start of each trial. Once the highlighting of letters begins, the green ellipse disappears. Each letter then transitions to white three times, and participants mentally count up to three with each highlighted letter. Participants are instructed to spell 10 four-letter words: HOME, WITH, WHAT, GOOD, YOUR, FROM, MUCH, THEM, 6-17, and 2345. These words are presented randomly, totaling 40 letters for spelling. At the trial's outset, the current word appears in white in the upper-left quadrant of the screen, with the target letter clearly indicated by a green ellipse for 1 second. Following a brief black screen interval, a 3-complete sequence (3CS) of flashing letters unfolds, lasting 4 seconds.

*Paradigm #5: P3005L (P300 for Five Letters Word).* The experimental setup for P3005L mirrors that of P3004L (refer to Fig. 2), with the sole alteration being the transition from four-letter to five-letter words. The chosen words for this task include ABOUT, BLACK, ENJOY, PRIZE, EQUAL, FALSE, HEAVY, EXACT, JUN88, and 13-59.

**Multimodal acquisition.** Three devices concurrently capture data from participants seated approximately 80 cm away from a 23-inch TFT monitor (refer to Fig. 3). The recording environment is a sound-attenuated, electromagnetically shielded chamber tailored for EEG acquisitions. To minimize head movements, participants are instructed to maintain their head on a chin rest while engaging in a BCI task. The chin rest, positioned at an angle of approximately 20°, is securely fixed to the table. Participants are afforded breaks between paradigms to relax and may move freely during these intervals. The monitor employed is a component of the Tobii TX300 Eye-Tracker (Tobii Technology AB, Stockholm, Sweden), with a resolution of 1920 × 1080 pixels and a refresh rate of 60 Hz.

*EEG.* Signals are acquired using a 65-channel Quik-cap interfaced with a SynAmps2 system connected to an amplifier (Compumedics, Neuroscan). Electrode placement follows the extended 10/20 system, with 62 EEG electrodes positioned accordingly. The reference electrode is situated on the right mastoid (M1), while the ground electrode is positioned on the forehead. A bipolar vertical EOG channel records potentials via two electrodes placed above and below the left eye. Additionally, two EMG electrodes capture electrical activity from the Levator Palpebrae Superioris and Orbicularis Oculi muscles around the right eye. These electrodes are taped to the middle of the upper and lower eyelids, respectively. Typically, EMG activity is sampled at a minimum of 1000 Hz, whereas EEG signals are commonly recorded at 250 Hz. As the electrodes transmit data to the same system, the protocol globally records the continuous signals at a 1000 Hz sampling frequency.

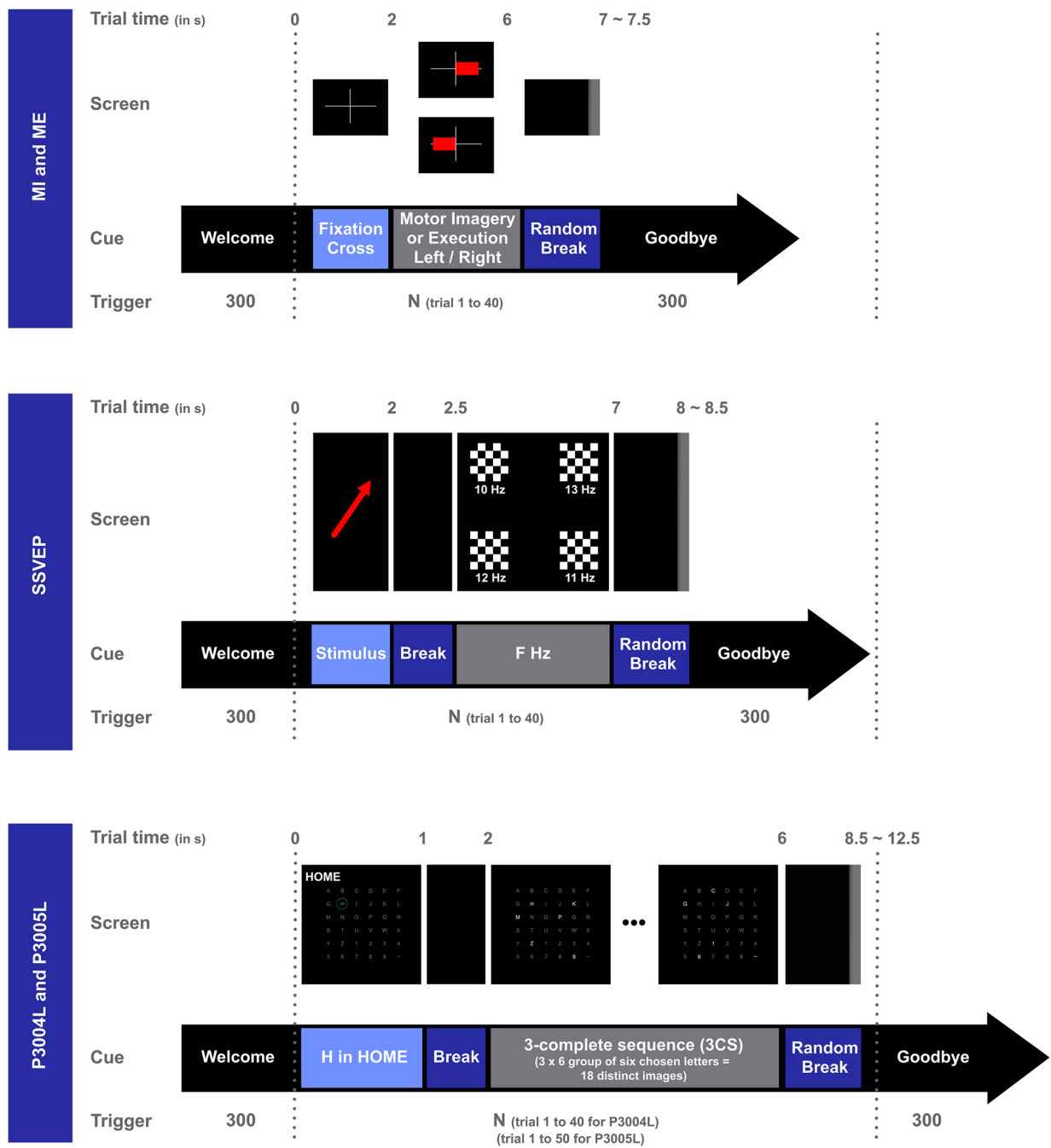

**Fig. 2** Chronology of the five paradigms: Motor Imagery and Motor Execution of a 1 Hz frequency hand grasping; multi-frequencies Steady-State Visual Evoked Potentials; P300 speller for four-letter and five-letter words.

*Eye-Tracker.* Gaze-related data is captured by an eye-tracker (Tobii TX300) operating at a sampling rate of 300 Hz. Eyeball position is determined through analysis of the pupil's movement. This process involves the use of an illuminator positioned at varying distances from the optical axis of the imaging device, resulting in alternating illumination and darkness of the pupil.

*High-Speed Camera.* A high-speed camera (Phantom Miro M310) records a single eye at a resolution of 320*240 pixels. This is the smallest mode to encompass the entire eye with a slight margin that can accommodate for minor head movements. The focus is primarily on the left eye, as the EOG electrodes attached to it are positioned farther from the eyelids than those on the right eye. This setup allows for the extraction of eyelid position from the video, operating under the assumption of symmetric blinking between both eyes.

*Multimodal Acquisition.* Cues are generated through the E-Prime software and projected onto the presentation screen. To overcome hardware limitations, the Tobii TX300's internal clock is used. Gaze-related data is captured approximately every 3.3 milliseconds, establishing a reference for sampling rates. Operating at a frequency comparable to Tobii's (300 fps) and equipped with a 10 GB internal memory, the Phantom high-speed camera enables a maximum recording duration of approximately seven minutes.

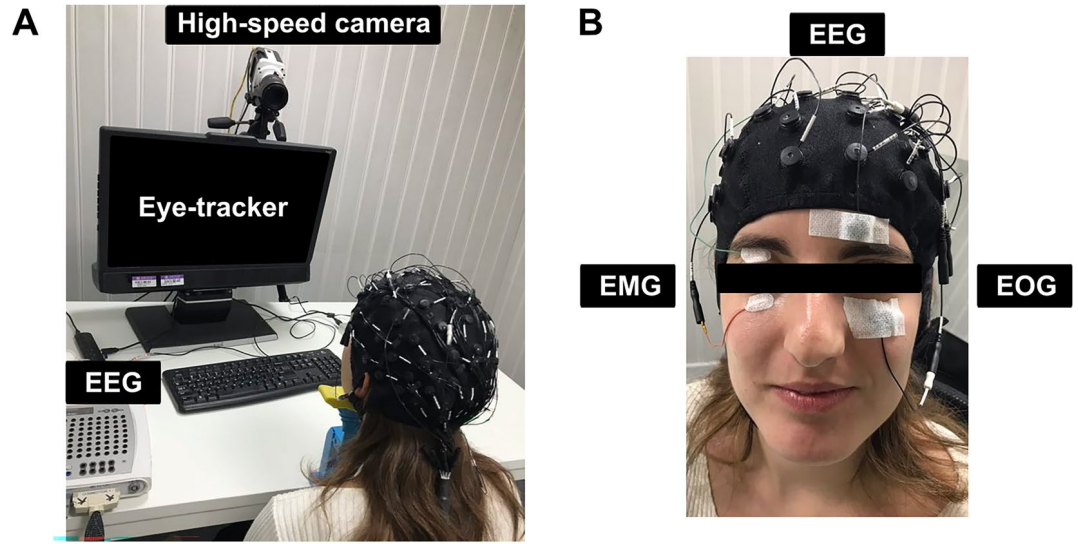

**Fig. 3** Data acquisition environment. Informed consent was obtained from the individual in the figure for the publication of the images.

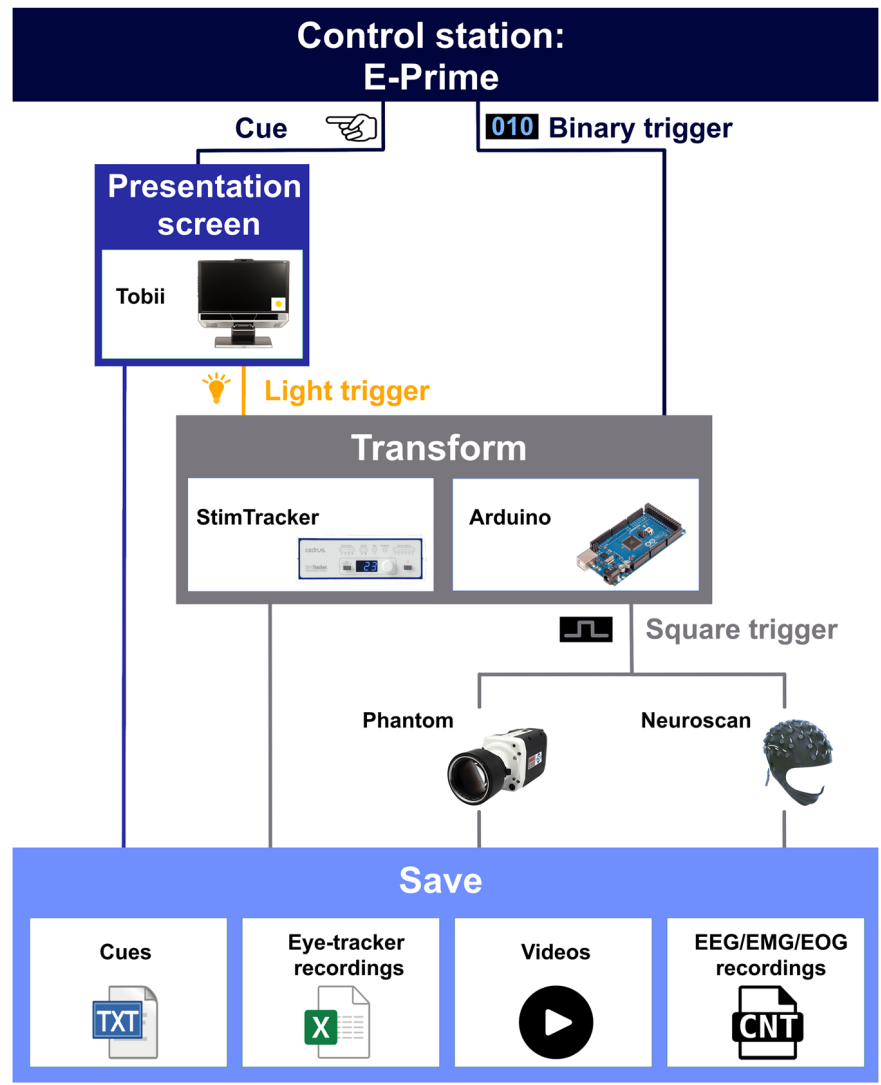

**Fig. 4** Experiment flowchart ensuring a common time base across the three devices: EEG, eye-tracking, and high-speed camera.

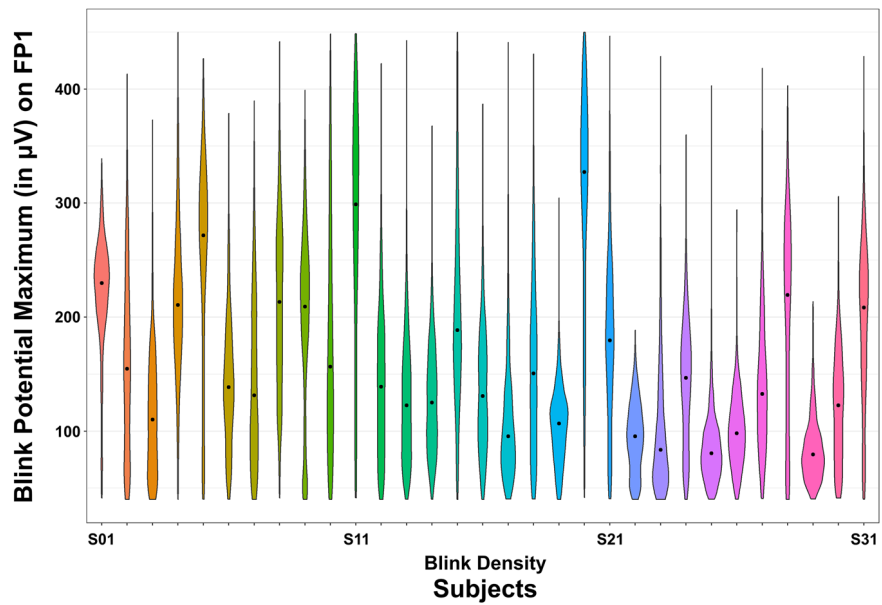

**Fig. 5** Violin/density plots showing the blink distribution on the frontopolar channel (FP1) for all subjects; the median is indicated as a black point.

To avoid any loss of data, the video sampling rate has been set at 150 fps. E-Prime emits a binary trigger every 6.6 milliseconds, received by an Arduino Nano (Atmega 328), which converts it into a square wave to regulate the Phantom camera shutter. The resulting videos are stored directly on the Phantom's internal memory and subsequently transferred to a computer using Phantom software following each task.

For synchronizing EEG and video recordings, the same triggers from E-Prime are also relayed to the Neuroscan software. As the Tobii TX300 cannot process triggers at such a high frequency, a light sensor detects a white rectangle on the screen, initiating a trigger via the Cedrus StimTracker. This ensures a common time base across all devices.

The four programs (E-Prime, Neuroscan, Phantom, and Tobii) operate on two separate computers to mitigate potential RAM-related problems. The first computer, acting as the control station, manages E-Prime, displaying cues on the presentation screen, and oversees eye-tracking data recording through the Tobii software. The second computer manages the two remaining softwares. At the experiment's onset, a Python code is launched to automatically initiate and stop recordings. Following each session, data synchronization is checked for consistency. The entire experimental process is illustrated in the flowchart depicted in Fig. 4.

A session commenced by randomly determining the sequence of paradigms. Participants responded to a brief questionnaire aimed at gauging their overall condition, including factors like sleepiness, coffee consumption, and hunger. Scheduled breaks were incorporated into the session structure to mitigate fatigue. During these intermissions, participants provided feedback on their alertness level, potential errors, and any external distractions encountered.

*Preprocessing.* Blinks are identified through the methodology outlined in[21], with their grand averages calculated for each session. The electrical activity recorded at each electrode is then evaluated against the median values from its neighboring electrodes and the Longest Common Subsequence (LCSS) is computed between these two trajectories. Channels are flagged as defective when their LCSS falls beneath a predefined threshold. It is recommended that any channel marked as defective in any session be excluded from the entire analysis. The electrodes identified as either malfunctioning or exhibiting bridging include: PO3, F1, POZ, OZ, F3, O2, P8, PO7, FC3, P7, and P4.

## Data Records
**Data privacy.** The dataset is anonymized in compliance with the informed consent protocol, and participants gave written permission for their data to be shared publicly. The EEG, eye-tracking, and high-speed video data are openly accessible, given the impossibility to identify individuals from this information without access to advanced equipment and a relevant database.

Although initial photographs from the experiment are kept confidential, facial landmarks extracted from these photos are accessible in the 'Info' folder at "Eye-BCI_multi_dataset"[22]. The Python script used for extracting these landmarks is also made available at https://github.com/QinXinlan/EEG-experiment-to-understand-differences-in-blinking/.

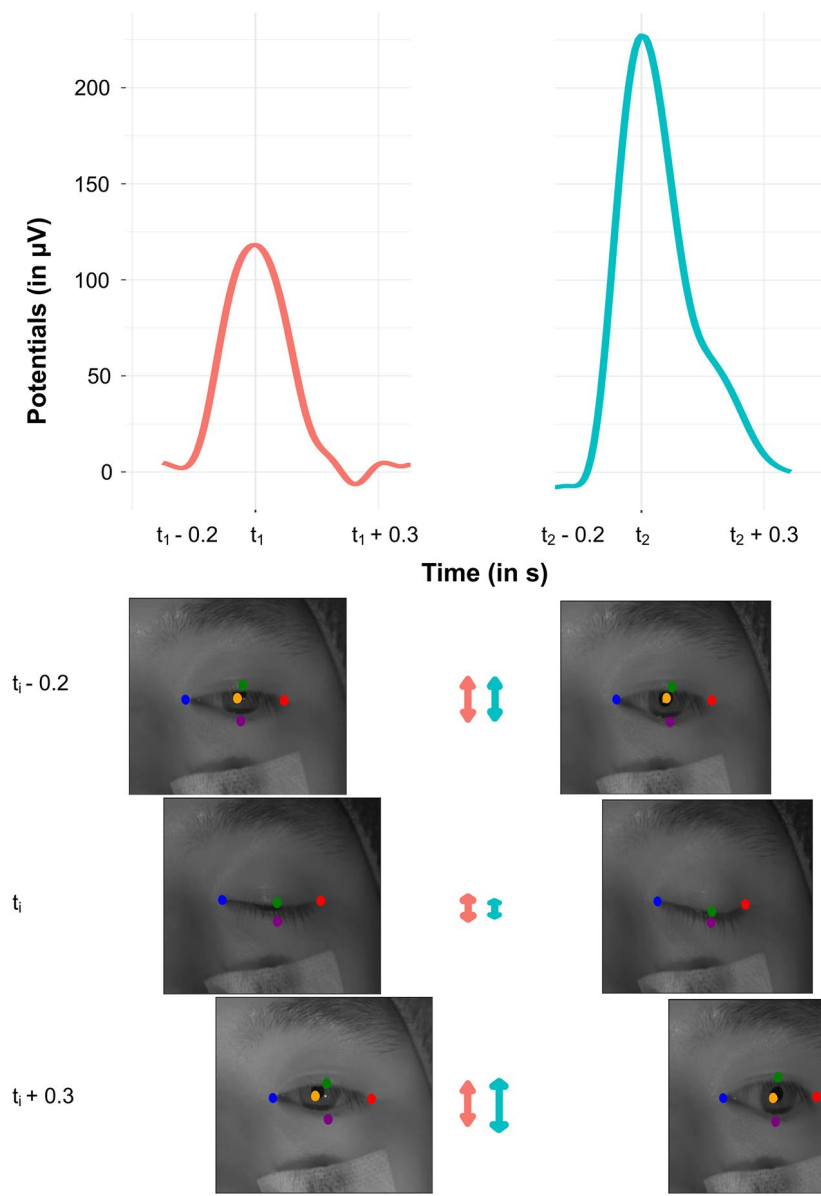

**Fig. 6** Comparison of blinks with different peak amplitudes (centered at $t_1$ and $t_2$); EEG blink signals extracted after fourth-order 10 Hz Butterworth low-pass filter on drift curve removed signals.

**Distribution for use.** The raw EEG, eye-tracking, and high-speed video data, along with the E-Prime cues and summaries of the questionnaires, is hosted in the Synapse project "Eye-BCI_multi_dataset"[22]. This dataset is shared under the CC0 License, available at https://creativecommons.org/public-domain/cc0/.

**Data structure organization.** The data files from the four software programs adhere to the following naming convention:

    NameParadigm-S-XX-Sess-Y

Where XX represents the participant ID ranging from 01 to 31, and Y indicates the session number, ranging from 1 to 3. The questionnaires are summarized for each session using the convention:

    S-XX-Sess-Y

Each participant's sessions are organized into respective folders, categorized by the four recording modalities. To facilitate comparative analysis, supplementary columns summarizing data from E-Prime (Trig and Cues columns) and the high-speed Phantom camera (PhanFrame, PhanTime, RelTime, RecordingTimestamp, and LocalTimeStamp) have been aggregated into the EEG files. Additionally, a Blinks column has been included, indicating either the absence (0) or the presence of a blink's peak (1). An illustrative row from an EEG file is presented in Table 2.

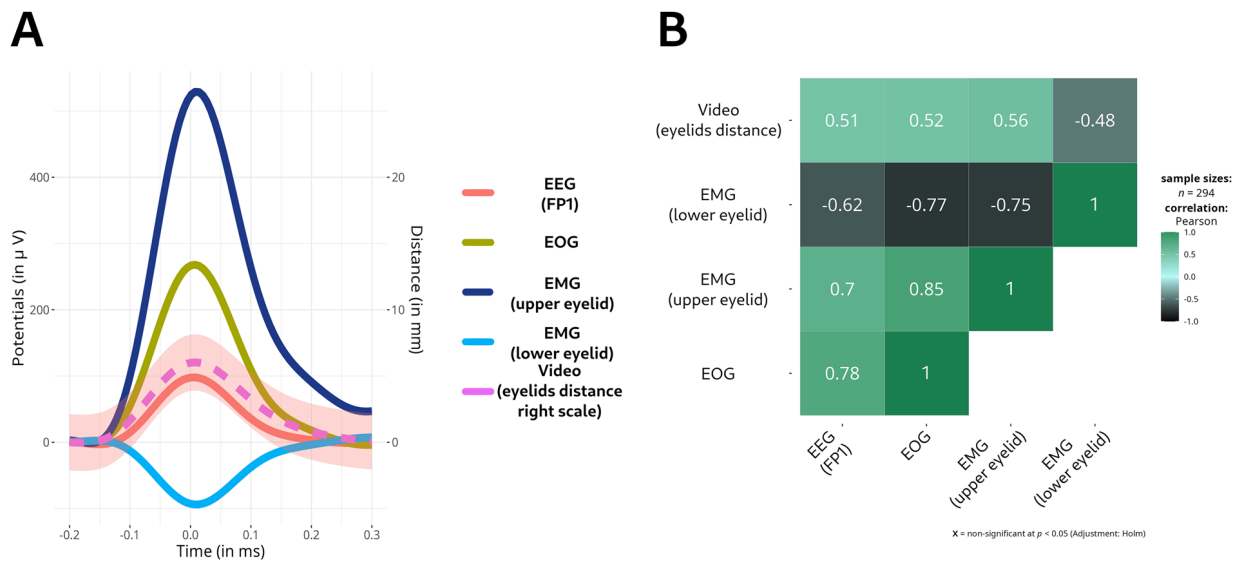

**A**

**B**

The Trig column indicates the trial number, ranging from 1 to 50 for the P3005L paradigm and from 1 to 40 for the other paradigms. The Cues column details the sequence within each trial, which differs across paradigms as depicted in Fig. 2. The PhanFrame (resp., PhanTime) column relates to the frame number (resp., timestamp) captured by the high-speed camera, information that can be found in the Xml file within the Phantom folder or directly on the video's lower portion.

For the Tobii eye tracker, key data include the gaze coordinates on the screen (GazePointLeftX, GazePointLeftY, GazePointRightX, GazePointRightY, GazePointX, GazePointY), the pupil positions within the camera frame (CamLeftX, CamLeftY, CamRightX, CamRightY), and the distance between the eye tracker and each eye (DistanceLeft, DistanceRight). Additionally, the pupil size for each eye is recorded (PupilLeft, PupilRight), along with validity codes (ValidityLeft, ValidityRight), which indicate the system's confidence in accurately identifying the left and right eyes. These measures are essential for ensuring accurate gaze and pupil tracking data.

**Missing data.** The dataset exhibited several instances of data absence across various modalities. In the electroencephalogram (EEG) recordings, only the initial trial of ME041 was missing (i.e., the ME task of subject S04 in session Sess01), constituting 0.01% of the total EEG trials.

In the video recordings, synchronization was not established for the trials in P3004L101 and ME181; consequently, approximate frame numbers per trial are provided for 80 trials, or 0.98% of the video data. Moreover, the Phantom camera initiated recording late, resulting in the absence of data from the beginning of each first trial for tasks ME011, MI011, ME091, MI091, P3004L091, SSVEP091, and ME093. This accounts for 0.09% of the video recordings being absent.

Regarding the eye-tracking data, recordings of visual stimuli were absent, presumably due to the incorrect attachment of the Cedrus light sensor to the display, for the entire first session of subject S03 (ME031, MI031, P3004L031, P3005L031, SSVEP031) and the whole first session of subject S06 (ME061, MI061, P3004L061, P3005L061, SSVEP061, MI241, and P3005L261). Additionally, incomplete recordings, likely due to a RAM issue or occasional detachment of the Cedrus light sensor, were observed for tasks P3004L023, ME052, MI132, MI232, MI241, P3005L261, and P3004L301. While eye-tracking data exist for all trials, the absence of a common time reference precludes comparison with EEG or video recordings, amounting to a total of 6.5% of eye-tracking trials. For all other tasks unaffected by sensor or RAM issues, an average of 0.11% (SD = 0.03%) of data is missing per trial.

## Technical Validation

The data quality can be evaluated through various methods, tailored to the specific features of interest. For instance, when analyzing blinks as the primary signal, their associated metrics can be examined to confirm data integrity. Alternatively, in research centered on BCI paradigms, the signal-to-noise ratio (SNR) or classification accuracy are dependable indicators of data quality. Recognized for their importance in BCI studies, these metrics highlight the dataset's reliability and technical robustness, crucial for ensuring overall validity.

Metrics related to eye movements, including blinks, pupil size variations, and gaze patterns, hold significant potential for various Human-Computer Interaction (HCI) applications. These metrics provide insights into user engagement, attention, and cognitive load, which are essential for adaptive interfaces. For instance, pupil size and gaze duration during tasks can indicate specific cognitive demands, creating a measurable basis for differentiating between BCI paradigms. This dataset can support the development of intelligent systems that dynamically adjust interface elements based on real-time cognitive feedback. Additionally, variations in gaze behavior and blink rates across paradigms present valuable opportunities for personalizing HCI systems to match user states, optimizing interaction efficiency, and enhancing user experience.

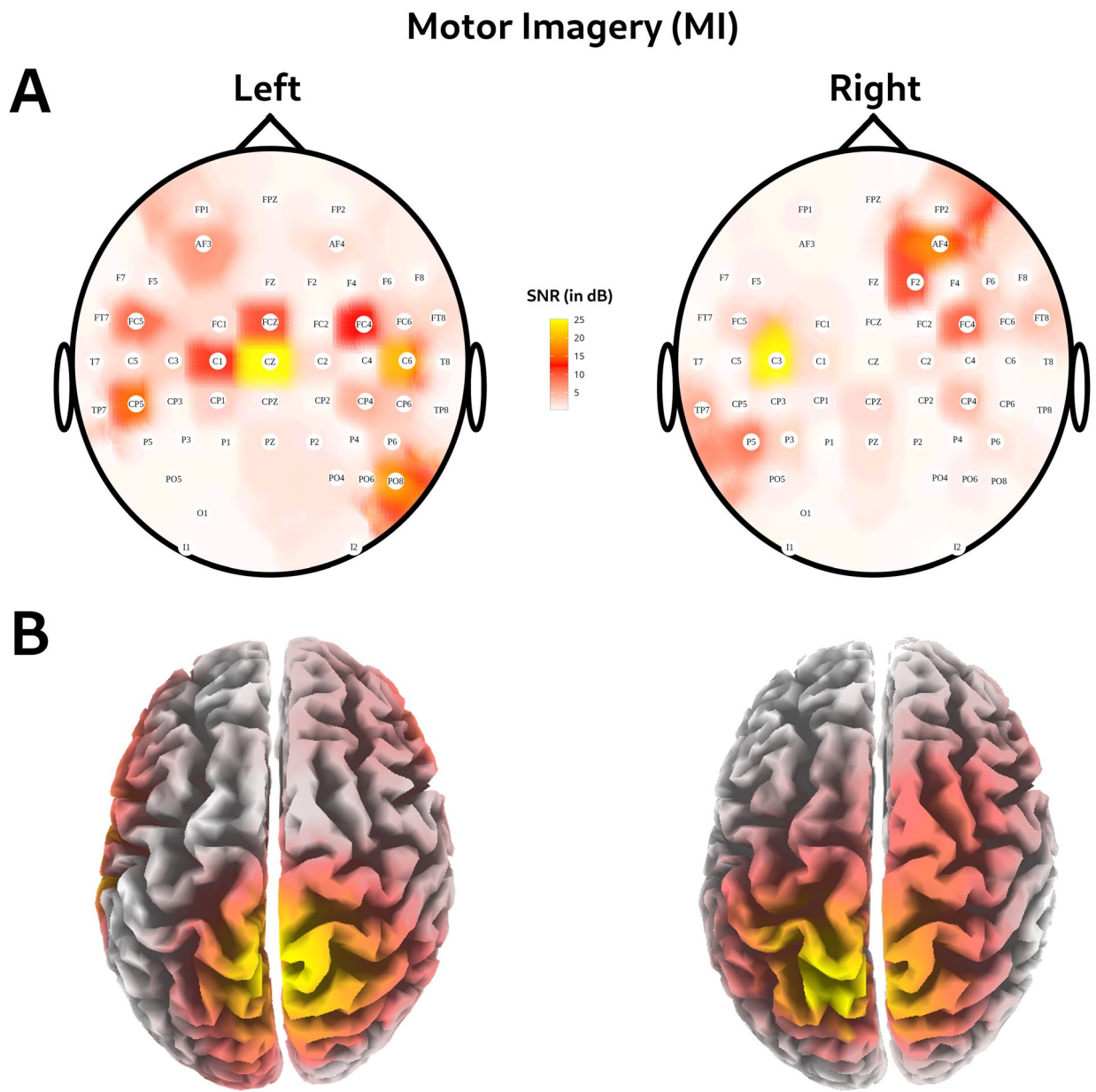

Fig. 8 (**A**) Segmented SNR topographies in the beta band (12.5-30 Hz) for both Left and Right MI (only the "good" electrodes are represented) and (**B**) Source localization plots.

In summary, the quality of the current dataset is validated using three key metrics: blink characteristics, data quality visualizations, including signal-to-noise ratio (SNR) plots, and classification accuracy. These preliminary analyses corroborate the dataset's robustness and consistency with prior findings, underscoring its potential for generating meaningful insights. This potential extends not only to blink-related studies but also to the different BCI paradigms under investigation. While these analyses provide a foundational assessment, the dataset offers substantial promise for further exploration in both BCI and HCI domains. By making it available, this resource is intended to support a wide range of research endeavors, facilitating the development of innovative algorithms and applications that harness eye movement and EEG data to advance adaptive, user-centered technologies. Additionally, it serves as a valuable benchmark for assessing and comparing the performance of new classification algorithms across various BCI paradigms.

**Blinks.**     Following a blink, tear fluid rapidly evaporates within 15 to 30 seconds, while the motion of the eyelids ensures the continuous lubrication of the cornea[23]. Although blink rates vary depending on the task at hand, they generally exceed the frequency required to maintain ocular moisture[24]. This phenomenon may be attributed to the need for multiple blinks to achieve a uniform distribution of tear fluid across the ocular surface[23]. An alternative hypothesis posits that increased blinking may function to momentarily disengage cognitive attention during demanding tasks[25]. Recent research has proposed that blinks may enhance visual sensitivity by generating luminance transients. These transients increase the power of retinal stimulation, particularly at low spatial frequencies, thereby contributing to improved contrast sensitivity during visual processing[26]. Additionally, despite the disruption to visual processing caused by blinks, they remain imperceptible and do not affect visual perception. This is attributed to neural mechanisms that maintain visual stability, including recurrent corticotha-lamic activity and suppression of visual transients[27].

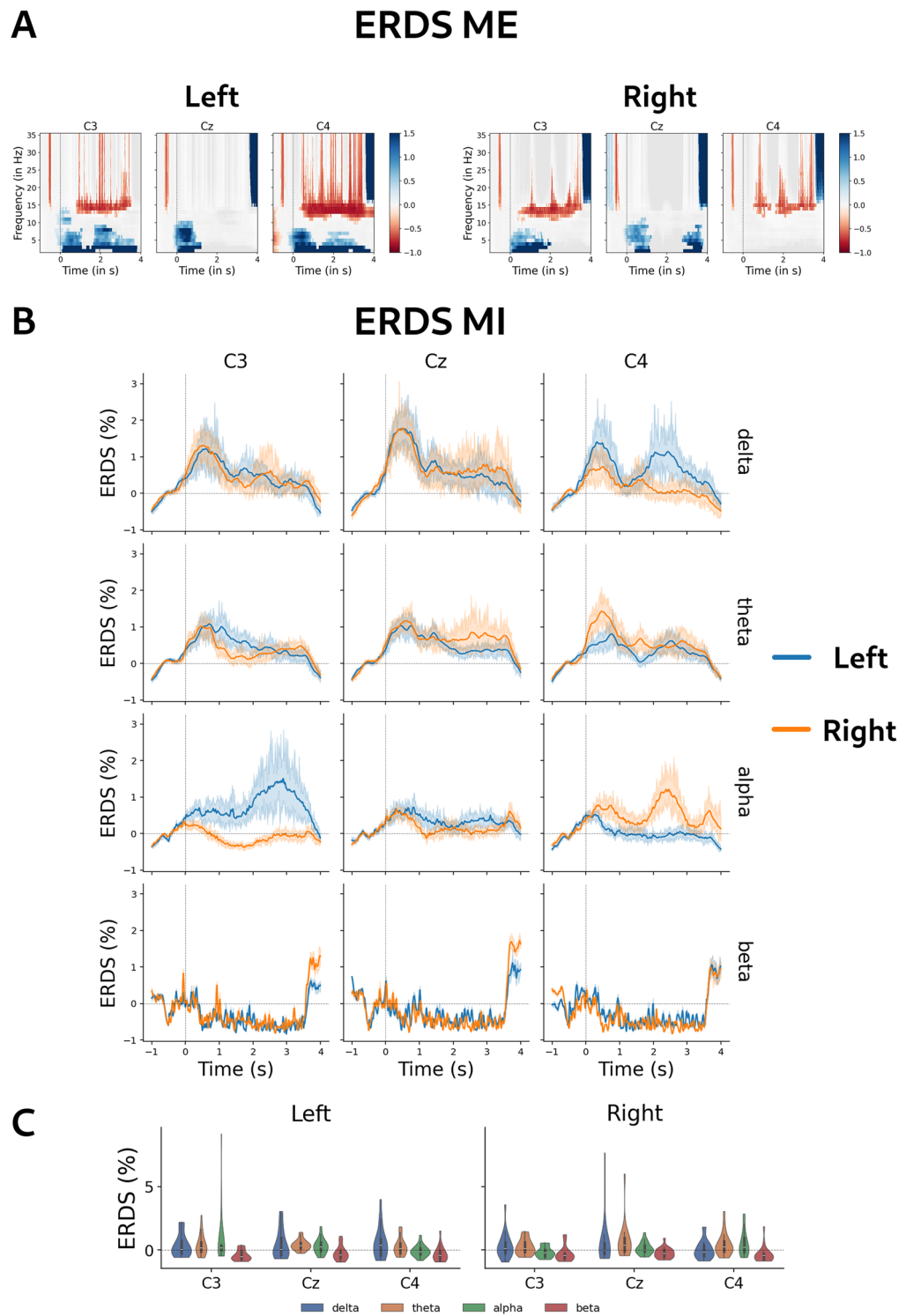

**Fig. 9** Illustrative example for a subject: (**A**) Time-frequency decomposition of Left and Right Event-Related Desynchronization/Synchronization (ERDS) for hand grasping Motor Execution (ME), (**B**) average ERDS with their corresponding confidence intervals on motor cortex-related electrodes, and (**C**) mean ERDS as a function of frequency band and Motor Imagery (MI) condition.

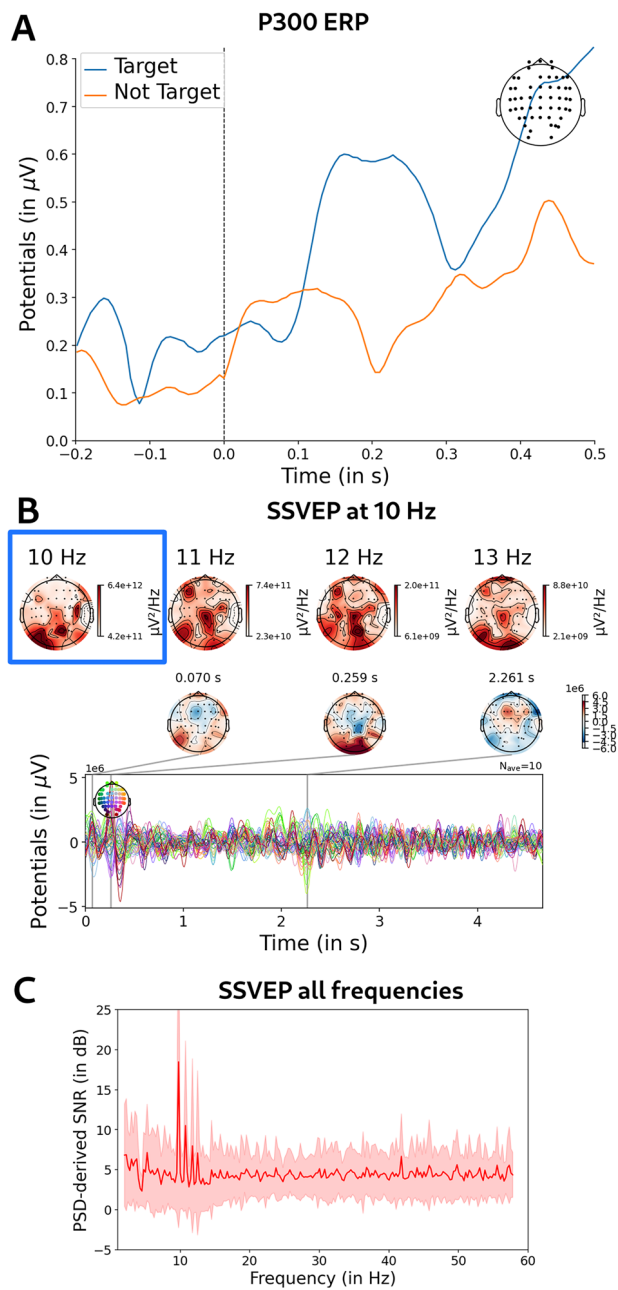

**Fig. 10** Illustrative example for a subject: (**A**) Event-Related Potentials (ERP) waveforms for target and non-target letters within the P300 speller, and Steady-State Visual Evoked Potentials (SSVEP) across temporal, spectral, and spatial domains at (**B**) a specific frequency and (**C**) across all frequencies.

Many spontaneous blinks exhibit incomplete closure, where the upper eyelid stops short of completely reaching the lower eyelid. The underlying cause of this phenomenon, whether it's to prevent unnecessary full closure or to minimize the duration of visual obstruction, remains uncertain. In any case, this strategy of incomplete closure contributes to significant intra-subject variability, as illustrated in Table 3 and Fig. 5.

For each frame of the video recordings, sub-images corresponding to critical eye regions — specifically, the inner and outer canthi, the midpoints of the upper and lower eyelids, and the pupil — are extracted using computer vision-based template matching. The relative positions of these key facial landmarks are then analyzed to detect blinks and to accurately determine their onset and offset. The code for this process, along with related resources, is available online at https://github.com/QinXinlan/EEG-experiment-to-understand-differences-in-blinking/. Additionally, visual inspection of these frames allows for verification of specific features that may contribute to data variability. This approach provides a deeper understanding of blink physiology while supporting improvements in the accuracy and efficiency of blink detection algorithms across modalities.

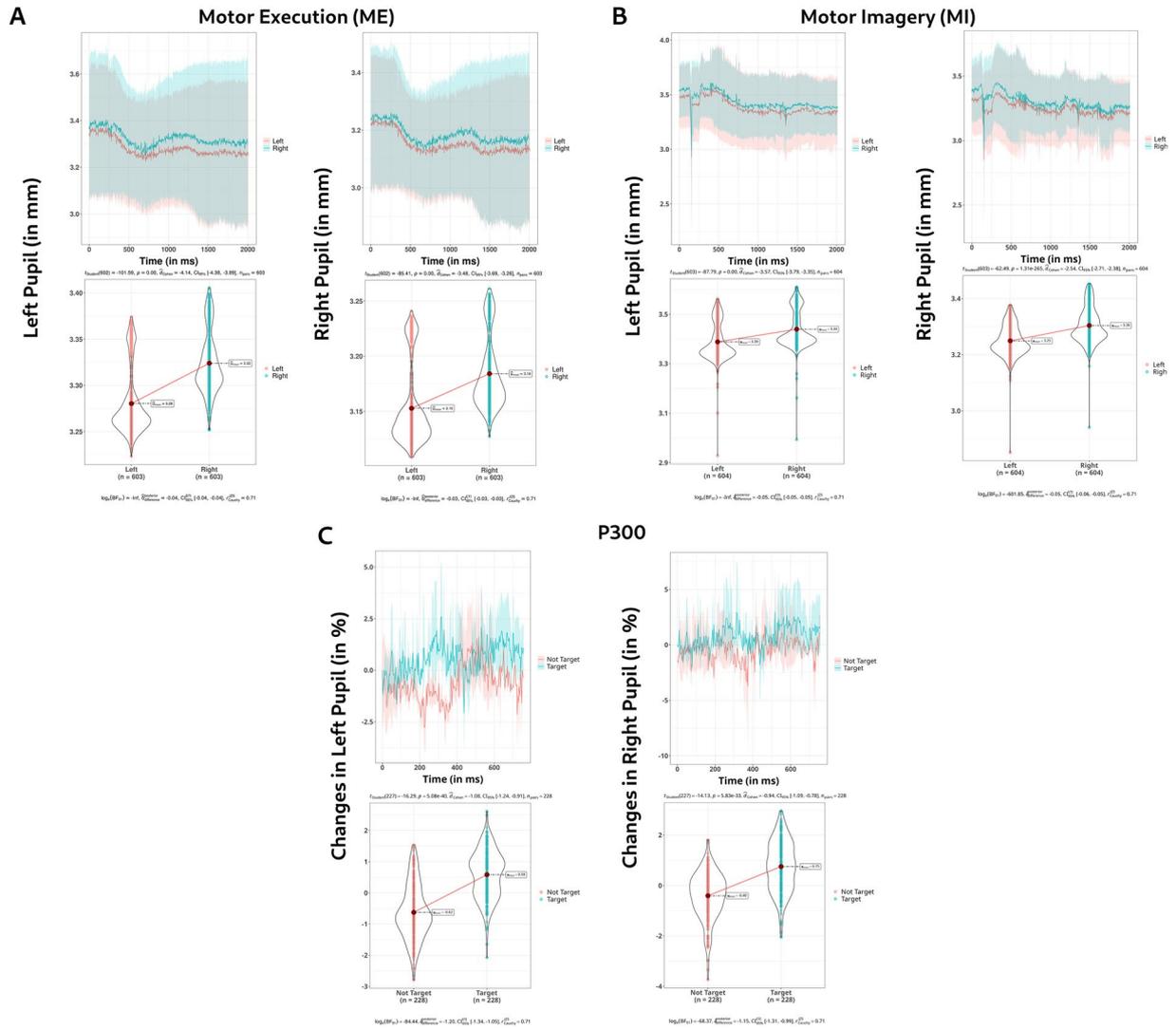

**Fig. 11** Illustrative example for a subject: Differences in Waveforms and Violin Plots of Pupil Sizes for Left and Right Eye During Target and Non-Target Conditions in (**A**) Motor Execution, (**B**) Motor Imagery, and (**C**) P300.

Blinks are also detected from EEG signals, using several criteria from amplitude to propagation[21]. The peak amplitude of a blink observed in frontopolar EEG channels indicates the overall movement of the eyelids, encompassing both closure and opening phases. Figure 6 presents two blinks with different peak amplitudes ($120\ \mu V$ and $220\ \mu V$), and their corresponding images from the high-speed camera at the onset and conclusion of the upper eyelid displacement.

This cross-modal example also validates the overall data quality by enabling comparative analysis between EEG and high-speed video recordings during blinks. By examining the correspondence between the timing and amplitudes of EEG signals and the observable physical movements captured by high-speed video, this dataset offers unique insights into blinks and their impact on EEG data.

The significant inter- and intra-subject variability observed in blink-related signals warrants an analysis of blink mean potentials on a per-subject basis. These mean potentials, representing the average signal associated with blink events, are examined across multiple sensors. A correlation matrix is then computed using data from all individual blinks[28] and is displayed alongside the mean potentials in Fig. 7, providing a visual representation of inter-modal consistency and variability. As shown, the frontopolar EEG channels, electrooculogram (EOG) electrodes (positioned above and below the left eye), and electromyogram (EMG) sensors (placed on the upper and lower eyelids of the right eye) record similar patterns, which are in alignment with the eyelid movement data captured in the video recordings.

**Signal-to-noise ratio (SNR) plots and data quality validation.** The literature extensively documents Event-Related Desynchronization (ERD), characterized by power reduction in the beta band during Motor Imagery (MI)[29–31]. Following the exclusion of defective channels and supplementary cleaning procedures[21], which are outside the scope of this study, EEG data are averaged per subject. Subsequently, Signal-to-Noise Ratio (SNR) plots in Fig. 8 are computed using the equation below:

$$SNR_T = \frac{\frac{1}{N}\sum_{i=1}^{N} \frac{1}{T} \int_0^T x_i^2(t)}{\frac{1}{N}\sum_{i=1}^{N} \frac{1}{T} \int_{t_1}^{t_1+T} x_i^2(t)} - 1 \tag{1}$$

| Subject | MI classification accuracy (in %) with confidence interval (CI) at confidence level = 95%; CI approximation = Clopper-Pearson | | |
|---|---|---|---|
| | ABCD | ICA | ASR |
| S01 | 98% [79%; 100%] | 88% [66%; 98%] | 93% [72%; 100%] |
| S02 | 89% [69%; 79%] | 86% [64%; 97%] | 88% [68%; 98%] |
| S03 | 100% [83%; 100%] | 84% [62%; 95%] | 80% [58%; 94%] |
| S04 | 89% [70%; 97%] | 77% [56%; 92%] | 83% [62%; 95%] |
| S05 | 99% [82%; 100%] | 79% [58%; 93%] | 80% [58%; 94%] |
| S06 | 95% [75%; 100%] | 60% [36%; 81%] | 100% [83%; 100%] |
| S07 | 97% [78%; 100%] | 78% [56%; 92%] | 85% [64%; 97%] |
| S08 | 96% [77%; 99%] | 72% [48%; 89%] | 84% [61%; 96%] |
| S09 | 87% [68%; 96%] | 74% [52%; 90%] | 93% [72%; 99%] |
| S10 | 91% [71%; 99%] | 66% [43%; 85%] | 78% [56%; 91%] |
| S11 | 85% [62%; 97%] | 75% [51%; 91%] | 65% [43%; 84%] |
| S12 | 98% [81%; 100%] | 83% [62%; 96%] | 83% [64%; 94%] |
| S13 | 92% [72%; 99%] | 80% [58%; 93%] | 89% [68%; 99%] |
| S14 | 98% [80%; 100%] | 76% [53%; 92%] | 80% [59%; 93%] |
| S15 | 88% [67%; 98%] | 83% [61%; 96%] | 89% [68%; 99%] |
| S16 | 95% [76%; 100%] | 93% [74%; 100%] | 86% [64%; 96%] |
| S17 | 95% [76%; 100%] | 89% [68%; 98%] | 80% [59%; 93%] |
| S18 | 97% [78%; 100%] | 88% [66%; 98%] | 77% [55%; 92%] |
| S19 | 90% [69%; 100%] | 65% [41%; 85%] | 85% [62%; 97%] |
| S20 | 95% [76%; 100%] | 58% [34%; 78%] | 95% [76%; 100%] |
| S21 | 70% [47%; 88%] | 75% [52%; 91%] | 85% [62%; 97%] |
| S22 | 100% [83%; 100%] | 83% [60%; 96%] | 75% [51%; 91%] |
| S23 | 94% [62%; 97%] | 73% [50%; 89%] | 88% [65%; 98%] |
| S24 | 100% [83%; 100%] | 78% [57%; 93%] | 95% [75%; 100%] |
| S25 | 93% [73%; 100%] | 70% [47%; 87%] | 90% [69%; 98%] |
| S26 | 73% [50%; 89%] | 75% [55%; 91%] | 68% [43%; 86%] |
| S27 | 95% [76%; 100%] | 93% [72%; 100%] | 78% [57%; 93%] |
| S28 | 100% [83%; 100%] | 95% [76%; 100%] | 93% [72%; 100%] |
| S29 | 100% [83%; 100%] | 68% [44%; 86%] | 88% [67%; 99%] |
| S30 | 100% [83%; 100%] | 78% [54%; 93%] | 98% [79%; 100%] |
| S31 | 100% [83%; 100%] | 68% [45%; 86%] | 88% [67%; 99%] |
| **Mean** | **93.81% [74.81%; 98.76%]** | **79.29% [57.41%; 92.89%]** | **84.05% [62.88%; 95.31%]** |

**Table 4.** Comparison of two-class Motor Imagery (MI) classification accuracies (fully automated pipeline) between the new ABCD and other existing methods, such as Independent Component Analysis (ICA) or Artifact Subspace Reconstruction (ASR), with corresponding confidence intervals (CI) at 95% confidence level.

The SNR plot illustrates the temporal average across electrodes. Additionally, source localization can be computed for each time point using eLORETA and visualized at the time relevant to the cortical activity linked with MI.

To demonstrate the data quality of the dataset, basic analyses of Event-Related Desynchronization/Synchronization (ERD/ERS) for Motor Execution (ME) and Motor Imagery (MI) tasks are presented. Figure 9 illustrates the temporal and spectral dynamics of power fluctuations within specific frequency bands, emphasizing the differences between left and right hand grasping. This figure includes time-frequency decompositions, averaged waveforms with corresponding confidence intervals, and the mean ERDS as a function of frequency band and imagery condition, serving as a foundational example of the dataset's reliability.

In addition, Event-Related Potentials (ERP) waveforms for target and non-target letters within the P300 speller paradigm are provided. Figure 10 further showcases the Steady-State Visual Evoked Potentials (SSVEP) across temporal, spectral, and spatial domains, offering an additional example of the dataset's overall integrity.

Additionally, analyses of the Tobii eye tracker data are included in Fig. 11, specifically focusing on the pupil sizes of PupilLeft and PupilRight for target and non-target letters in the P300 speller paradigm and left versus right hand during ME and MI tasks. These illustrative analyses offer further insights into the dataset's quality and usability[28].

**Accuracy.** Classification accuracy serves as an implicit data quality metric in most BCI research. Its high values not only demonstrate an algorithm's efficacy in distinguishing between various classes or categories but also signify the quality of the utilized data for model training and testing. Furthermore, classification accuracy is easily interpretable and intuitive. This enhances comprehension and communication regarding a dataset's reliability and suitability for any intended analysis. That is why we assume that employing classification accuracy as a data quality metric provides a robust and practical means of assessing the dataset's effectiveness and reliability for BCI paradigms.

The accuracy-based data quality assessment depends on two primary factors: the quality of the raw recordings and the efficacy of blink correction. First, the quality of the EEG recordings is crucial, as suboptimal data can undermine the classification process ("garbage in, garbage out"). Second, the effectiveness of blink correction varies across algorithms, with ICA[32], ASR[33], and the original ABCD method outlined in a prior publication[21], each providing distinct approaches to artifact mitigation. The resulting classification accuracy reflects a combination of recording quality as well as blink detection and correction accuracy. For the two-class Motor Imagery (MI) paradigm, the ABCD method achieves a classification accuracy of 94%, as presented in Table 4, underscoring the dataset's high quality and suitability for BCI applications. The observed differences between the algorithms are likely due to the presence of blinks in approximately 20% of the trials, as shown in Table 3. While not all signals of interest are affected by blinks, it is reasonable to assume that a significant portion is, which may account for some of the observed discrepancies. Comparable accuracies were observed across additional paradigms, though these are beyond the scope of the current paper.

## Usage Notes

Code for loading the data in Matlab, Python, and R is available at https://github.com/QinXinlan/EEG-experiment-to-understand-differences-in-blinking/.

## Code availability

Comprehensive technical insights regarding the experiment, along with detailed explanations of the overall setup and computer codes necessary for replication, are accessible online at https://github.com/QinXinlan/EEG-experiment-to-understand-differences-in-blinking/.

## References

1. Tan, D. S. & Nijholt, A. (eds.). *Brain-Computer Interfaces and Human-Computer Interaction*, chap. 1. Brain-Computer Interfaces. Human-Computer Interaction Series (Springer, London, 2010).
2. Mridha, M. F. *et al.* Brain-Computer Interface: Advancement and Challenges. *Sensors (Basel)* **21**, 5746, https://doi.org/10.3390/s21175746 (2021).
3. Olejniczak, P. Neurophysiologic Basis of EEG. *Journal of Clinical Neurophysiology* **23**, 186–189, https://doi.org/10.1097/01.wnp.0000220079.61973.6c (2006).
4. Urigüen, J. A. & Garcia-Zapirain, B. EEG artifact removal—state-of-the-art and guidelines. *Journal of Neural Engineering* **12**, https://doi.org/10.1088/1741-2560/12/3/031001 (2015).
5. Tatum, W. O., Dworetzky, B. A. & Schomer, D. L. Artifact and Recording Concepts in EEG. *Journal of Clinical Neurophysiology* **28**, 252–263, https://doi.org/10.1097/WNP.0b013e31821c3c93 (2011).
6. Królak, A. & Strumiłło, P. Eye-blink detection system for human–computer interaction. *Universal Access in the Information Society* **11**, 409–419, https://doi.org/10.1007/s10209-011-0256-6 (2012).
7. Nakano, T., Yamamoto, Y., Kitajo, K., Takahashi, T. & Kitazawa, S. Synchronization of spontaneous eyeblinks while viewing video stories. *Proceedings of the Royal Society B: Biological Sciences* **276**, 3635–3644, https://doi.org/10.1098/rspb.2009.0828 (2009).
8. Jia, Y. & Tyler, C. W. Measurement of saccadic eye movements by electrooculography for simultaneous EEG recording. *Behavior Research Methods* **51**, 2139–2151, https://doi.org/10.3758/s13428-019-01280-8 (2019).
9. Daly, I., Matran-Fernandez, A., Valeriani, D., Lebedev, M. & Kübler, A. Editorial: Datasets for Brain-Computer Interface Applications. *Frontiers in Neuroscience* **15**, 732165, https://doi.org/10.3389/fnins.2021.732165 (2021).
10. Agarwal, M. & Sivakumar, R. Blink: A Fully Automated Unsupervised Algorithm for Eye-Blink Detection in EEG Signals. In *2019 57th Annual Allerton Conference on Communication, Control, and Computing (Allerton)*, 1113–1121, https://doi.org/10.1109/ALLERTON.2019.8919795 (2019).
11. Brunner, C., Leeb, R., Müller-Putz, G., Schlögl, A. & Pfurtscheller, G. BCI Competition 2008–Graz data set A. *Institute for knowledge discovery (laboratory of brain-computer interfaces), Graz University of Technology* **16**, 1–6, https://doi.org/10.21227/katb-zv89 (2008).
12. Arvaneh, M., Guan, C., Ang, K. K. & Quek, C. Optimizing the Channel Selection and Classification Accuracy in EEG-based BCI. *IEEE transactions on bio-medical engineering* **58**, 1865–1873, https://doi.org/10.1109/TBME.2011.2131142 (2011).
13. Ellis, P. D. *The Essential Guide to Effect Sizes: Statistical Power, Meta-Analysis, and the Interpretation of Research Results* (Cambridge University Press, Cambridge, 2010).
14. Guttmann-Flury, E., Sheng, X., Zhang, D. & Zhu, X. A Priori Sample Size Determination for the Number of Subjects in an EEG Experiment. In *2019 41st Annual International Conference of the IEEE Engineering in Medicine and Biology Society (EMBC)*, 5180–5183, https://doi.org/10.1109/EMBC.2019.8857482 (2019).
15. Oldfield, R. C. The assessment and analysis of handedness: The Edinburgh inventory. *Neuropsychologia* **9**, 97–113, https://doi.org/10.1016/0028-3932(71)90067-4 (1971).
16. Reeves, R. R., Struve, F. A. & Patrick, G. A Comprehensive Questionnaire for Subjects Undergoing Quantitative Research EEGs. *Clinical Electroencephalography* **29**, 67–72, https://doi.org/10.1177/155005949802900204 (1998).
17. Sagonas, C., Antonakos, E., Tzimiropoulos, G., Zafeiriou, S. & Pantic, M. 300 Faces In-The-Wild Challenge: database and results. *Image and Vision Computing* **47**, 3–18, https://doi.org/10.1016/j.imavis.2016.01.002 (2016).
18. Farwell, L. A. & Donchin, E. Talking off the top of your head: toward a mental prosthesis utilizing event-related brain potentials. *Electroencephalography and Clinical Neurophysiology* **70**, 510–523, https://doi.org/10.1016/0013-4694(88)90149-6 (1988).
19. Townsend, G. *et al.* A novel P300-based brain-computer interface stimulus presentation paradigm: Moving beyond rows and columns. *Clinical Neurophysiology* **121**, 1109–1120, https://doi.org/10.1016/j.clinph.2010.01.030 (2010).
20. Cecotti, H. & Rivet, B. One step beyond rows and columns flashes in the P300 speller: a theoretical description. *International Journal of bioelectromagnetism* **13**, 39–41 (2010). Publisher: International Society for Bioelectromagnetism.

21. Guttmann-Flury, E., Sheng, X., Zhang, D. & Zhu, X. A new algorithm for blink correction adaptive to inter- and intra-subject variability. *Computers in Biology and Medicine* **114**, 103442, https://doi.org/10.1016/j.compbiomed.2019.103442 (2019).
22. Guttmann-Flury, E., Sheng, X., Zhang, D. & Zhu, X. Eye-BCI multimodal dataset. *Synapse* https://doi.org/10.7303/syn64005218 (2024).
23. Doane, M. G. Interaction of Eyelids and Tears in Corneal Wetting and the Dynamics of the Normal Human Eyeblink. *American Journal of Ophthalmology* **89**, 507–516, https://doi.org/10.1016/0002-9394(80)90058-6 (1980).
24. Bentivoglio, A. R. *et al*. Analysis of Blink Rate Patterns in Normal Subjects. *Movement Disorders* **12**, 1028–1034, https://doi.org/10.1002/mds.870120629 (1997).
25. Nakano, T. Blink-related dynamic switching between internal and external orienting networks while viewing videos. *Neuroscience Research* **96**, 54–58, https://doi.org/10.1016/j.neures.2015.02.010 (2015).
26. Yang, B., Intoy, J. & Rucci, M. Eye blinks as a visual processing stage. *Proceedings of the National Academy of Sciences* **121**, e2310291121, https://doi.org/10.1073/pnas.2310291121 (2024).
27. Willett, S. M., Maenner, S. K. & Mayo, J. P. The perceptual consequences and neurophysiology of eye blinks. *Frontiers in Systems Neuroscience* **17**, 1242654, https://doi.org/10.3389/fnsys.2023.1242654 (2023).
28. Patil, I. Visualizations with statistical details: The 'ggstatsplot' approach. *Journal of Open Source Software* **6**, 3167, https://doi.org/10.21105/joss.03167 (2021).
29. Kraeutner, S., Gionfriddo, A., Bardouille, T. & Boe, S. Motor imagery-based brain activity parallels that of motor execution: Evidence from magnetic source imaging of cortical oscillations. *Brain Research* **1588**, 81–91, https://doi.org/10.1016/j.brainres.2014.09.001 (2014).
30. Nam, C. S., Jeon, Y., Kim, Y.-J., Lee, I. & Park, K. Movement imagery-related lateralization of event-related (de)synchronization (ERD/ERS): Motor-imagery duration effects. *Clinical Neurophysiology* **122**, 567–577, https://doi.org/10.1016/j.clinph.2010.08.002 (2011).
31. Burianová, H. *et al*. Multimodal functional imaging of motor imagery using a novel paradigm. *NeuroImage* **71**, 50–58, https://doi.org/10.1016/j.neuroimage.2013.01.001 (2013).
32. Makeig, S., Bell, A., Jung, T.-P. & Sejnowski, T. J. Independent Component Analysis of Electroencephalographic Data. In *Advances in Neural Information Processing Systems*, vol. 8 (MIT Press, 1995).
33. Kothe, C. A. E. & Jung, T.-P. Artifact removal techniques with signal reconstruction (2016). US Patent App. 14/895,440.

## Acknowledgements

This work is supported by the National Natural Science Foundation of China (Grant No. 91948302).

## Author contributions

E.G.F. participated in designing the experiment, collecting data, programming software, validating data, and preparing the manuscript. X.S. contributed to the experiment's design and manuscript preparation. X.Z. supervised the project and edited the manuscript.

## Competing interests

The authors declare no competing interests.

## Additional information

**Correspondence** and requests for materials should be addressed to E.G.-F.

**Reprints and permissions information** is available at www.nature.com/reprints.

**Publisher's note** Springer Nature remains neutral with regard to jurisdictional claims in published maps and institutional affiliations.